\documentclass[tightenlines,nofootinbib,showpacs]{revtex4}

\usepackage{graphicx}

\newcommand{\be}{\begin{equation}}
\newcommand{\ee}{\end{equation}}
\newcommand{\ba}{\begin{eqnarray}}
\newcommand{\ea}{\end{eqnarray}}
\newcommand{\ban}{\begin{eqnarray*}}
\newcommand{\ean}{\end{eqnarray*}}

\begin{document}
\draft

\title{Correlation of transverse momentum and multiplicity \\
in a superposition model of nucleus--nucleus collisions}

\author{Stanis\l aw Mr\'owczy\'nski\\[2mm]}

\address{Institute of Physics, \'Swi\c etokrzyska Academy,
ul. \'Swi\c etokrzyska 15, PL - 25-406 Kielce, Poland \\
and So\l tan Institute for Nuclear Studies,
ul. Ho\.za 69, PL - 00-681 Warsaw, Poland}

\date{28-th March 2006}

\begin{abstract}

In p--p collisions the average transverse momentum is known to be 
correlated with the multiplicity of produced particles. The correlation 
is shown to survive in a superposition model of nucleus--nucleus 
collisions. When properly parameterized, the correlation strength 
appears to be independent of the collision centrality  - it is the 
same in p--p and central A--A collisions. However, the correlation
is strongly suppressed by the centrality fluctuations.

\end{abstract}

\vspace{0.3cm}


%25.75.-q Relativistic heavy-ion collisions 
%25.75.Gz Particle correlations

%\vspace{0.5cm}

\maketitle

%%%%%%%%%%%%%%%%%%%%%%%%%%%%%%%%%%%%%%%%%%%%%%%%%%%%%%%%%%%%
\section{Introduction}
%%%%%%%%%%%%%%%%%%%%%%%%%%%%%%%%%%%%%%%%%%%%%%%%%%%%%%%%%%%%

The transverse momentum of particles produced in p--p collisions 
is correlated with the particle's multiplicity. At the SPS energy 
the correlation, which is directly observed experimentally 
\cite{Kafka:1976py,Anticic:2003fd}, is most probably of simple 
kinematical origin - when the multiplicity $N$ of produced particles 
at fixed collision energy grows, there is less and less energy 
to be distributed among transverse degrees of freedom of produced 
particles. Consequently, the average transverse momentum 
$\langle p_T \rangle_{(N)}$ decreases when $N$ grows. The
dependence of $\langle p_T \rangle_{(N)}$ on $N$ is approximately 
linear \cite{Kafka:1976py,Anticic:2003fd}.

The correlation $p_T$~vs.~$N$ seems to dominate the transverse 
momentum fluctuations which have been experimentally studied 
\cite{Anticic:2003fd} in nucleus--nucleus (A--A) collisions at 
SPS energy. The fluctuations have been quantified by the measure 
$\Phi(p_T)$ \cite{Gazdzicki:ri} and it has been shown 
\cite{Mrowczynski:2004cg} that the non-monotonic behavior of 
$\Phi(p_T)$ as a function of collision centrality can be explained 
by the observed non-monotonic multiplicity fluctuations 
\cite{Gazdzicki:2004ef,Rybczynski:2004yw} combined with the 
correlation $p_T$~vs.~$N$. It contradicts an expectation 
that the correlation, which is well seen in p--p interactions, 
is washed out in A--A collisions. To clarify the situation 
the NA49 Collaboration has undertaken an effort to directly 
measure the correlation $p_T$~vs.~$N$ in A--A collisions
at precisely determined number of projectile participants.

An analysis of the experimental data, which is underway, is 
performed for p--p and Pb--Pb collisions at 158 AGeV in the 
forward rapidity window $[4.0,5.5]$ (the center-of-mass rapidity 
is 2.92). The collision centrality is determined on the 
event-by-event basis, using the zero-degree calorimeter which 
measures the energy deposited by remnants of the projectile
nucleus. The energy is recalculated into the number of spectator 
nucleons coming from the projectile. The experimental method 
was worked out in the course of multiplicity fluctuation 
measurements performed at precisely fixed collision 
centralities \cite{Gazdzicki:2004ef,Rybczynski:2004yw}.

The motivation of our analysis is twofold. The first goal, which 
is oriented towards experimentalist's needs, is to check 
how a finite calorimeter resolution, which hinders the centrality
determination, influences the observed correlation. The second
goal is to formulate predictions within simple models of 
nucleus--nucleus collisions as a reference for the experimental 
data. For these purposes we discuss the correlation $p_T$~vs.~$N$ 
in A-A collisions, assuming that the particles produced in such 
a collision originate from several identical sources which are 
independent from each other. A particular realization is the model 
where nucleus--nucleus collision is a superposition of independent 
nucleon--nucleon interactions. Then, a single nucleon--nucleon 
interaction is a single source of particles. More realistic 
is the Wounded Nucleon Model \cite{Bialas:1976ed} where the wounded 
nucleons - those which interact at least once - contribute independently 
from each other to the final multiplicity. Then, there are two wounded
nucleons in a nucleon--nucleon interaction, and thus, there are two 
particle's sources. Since the particles in the forward domain are 
analyzed by the NA49 Collaboration, only the wounded nucleons form 
a projectile are expected to matter, and then their number equals the
number of particle's sources. 

Our results obtained within the approach of independent sources 
can be interpreted, referring to one of three categories of models 
of nucleus-nucleus collisions introduced in \cite{Gazdzicki:2005rr}. 
There are {\it transparency models} when the projectile (target) 
participants, which are excited in the course of interaction, mostly 
contribute to the particle production in the forward (backward) hemisphere. 
One deals with (rather unrealistic) {\it reflection models} when 
the projectile (target) participants mostly contribute to the 
backward (forward) hemisphere. Finally, there are {\it mixing models}
when the projectile and target participants equally contribute 
to the backward and forward hemispheres. The latter models were
shown \cite{Gazdzicki:2005rr,Konchakovski:2005hq} to properly 
describe (large) multiplicity fluctuations of particles produced 
in the forward hemisphere at fixed number of projectile 
participants \cite{Gazdzicki:2004ef,Rybczynski:2004yw}, as the 
target participants of fluctuating number contribute to the 
forward hemisphere. The transparency models highly underestimate
the multiplicity fluctuations while the reflection models strongly
overestimate them.

Because of various possible interpretations of our results, we use
in Secs.~\ref{1-source}-\ref{v-sources} the term {\it source} not 
referring to a specific model. We first discuss the correlation 
$p_T$~vs.~$N$ for particles coming from a single source 
(Sec.~\ref{1-source}), and then, we consider particles which originate 
from several identical and independent sources (Sec.~\ref{k-sources}). 
In Sec.~\ref{v-sources} a situation with varying number of particle 
sources is discussed. Our results are interpreted in two models in 
Sec.~\ref{models} where specific predictions are presented. To 
simplify the notation, the transverse momentum is denoted from now 
on as $p$ not as $p_T$ or $p_\perp$. The word `transverse' is also 
mostly skipped in the text.

%%%%%%%%%%%%%%%%%%%%%%%%%%%%%%%%%%%%%%%%%%%%%%%%%%%%%%%%%%%%
\section{Single source}
\label{1-source}
%%%%%%%%%%%%%%%%%%%%%%%%%%%%%%%%%%%%%%%%%%%%%%%%%%%%%%%%%%%%

$P_N(p_1,p_2, \dots ,p_N)$ is the probability density that
$N$ particles with momenta $p_1,\;p_2,\; \dots p_N$ are produced 
by a single source. It is normalized as 
\be
\label{normal1}
\sum_N \int dp_1 dp_2 \dots dp_N P_N(p_1,p_2, \dots ,p_N) = 1\;.
\ee
The multiplicity distribution ${\cal P}_N$ is given by
\ban
{\cal P}_N \buildrel \rm def \over =
 \int dp_1 dp_2 \dots dp_N P_N(p_1,p_2, \dots ,p_N) \;.
\ean

The momentum distribution of a single particle in the events 
of the multiplicity $N$ is defined as
\ba
\label{singel-distri}
P_{(N)}(p) \buildrel \rm def \over =  \frac{1}{{\cal P}_N} \frac{1}{N} 
&\bigg[& \int dp_2 \dots dp_N P_{(N)}(p,p_2, \dots ,p_N) 
+ \int dp_1 dp_3 \dots dp_N P_{(N)}(p_1,p,p_3 \dots ,p_N) 
\\ \nonumber
&+& \cdots + \int dp_1 dp_2 \dots dp_{N-1} P_{(N)}(p_1,p_2 \dots ,p_{N-1},p)  
\bigg] \;,
\ea
and it is also normalized 
\ban
\int dp P_{(N)}(p) = 1\;.
\ean

The single particle average momentum in the events with $N$
particles, which is denoted as $\langle p \rangle_{(N)}$, is 
correlated with $N$, and this correlation has to be parameterized 
for our further considerations. According to the data 
\cite{Kafka:1976py,Anticic:2003fd}, the dependence of 
$\langle p \rangle_{(N)}$ on $N$ is approximately 
linear, and thus we write  
\be 
\label{pT-N-corr1}
\langle p \rangle_{(N)} = a 
+ b \bigg(1 - \frac{N}{\langle N \rangle} \bigg) \;.
\ee
We note that the parameter $a$ only approximately equals 
the inclusive average momentum $\langle p \rangle$ which is 
\be
\label{ave-inclusiv}
\langle p \rangle \buildrel \rm def \over =  
\frac{1}{\langle N \rangle}
\sum_N N {\cal P}_N \langle p \rangle_{(N)} \;,
\ee
with
\ban
\langle N \rangle \buildrel \rm def \over =  
\sum_N N {\cal P}_N \;.
\ean
The parameterization (\ref{pT-N-corr1}) gives
\ban
\langle p \rangle = a + b 
\frac{\langle N \rangle^2 - \langle N^2 \rangle}{\langle N \rangle^2} \;,
\ean
which for the case of poissonian multiplicity distribution is
\ban
\langle p \rangle = a - 
\frac{b}{\langle N \rangle} \;.
\ean

The multiplicity distribution of charged particles produced in 
high-energy proton-proton collisions is known not to be poissonian. 
First of all, the total number of charged particles is always even 
due to the charge conservation. The multiplicity distribution of 
positive or negative particles is not poissonian as well - the 
variance does not grow linearly with $\langle N \rangle$ but it 
follows the so-called Wr\'oblewski formula \cite{Wroblewski:1973tn} 
{\it i.e.} the variance is a quadratic function of $\langle N \rangle$. 
However, for the collision energies of the SPS domain, which are of 
interest here, the Poisson distribution provides a reasonable 
approximation. The multiplicity distribution of not only positive 
or negative but of all charge particles is poissonian with high 
accuracy when the particles are registered in a limited acceptance. 
Therefore, the multiplicity distribution, which is further used 
in our calculation, is assumed to be of the Poisson form.

%%%%%%%%%%%%%%%%%%%%%%%%%%%%%%%%%%%%%%%%%%%%%%%%%%%%%%%%%%%%
\section{Several sources}
\label{k-sources}
%%%%%%%%%%%%%%%%%%%%%%%%%%%%%%%%%%%%%%%%%%%%%%%%%%%%%%%%%%%%

The momentum distribution of $N$ particles coming from $k$ 
identical and independent sources equals
\ba
\label{k-source-distri}
P_N^k(p_1,p_2, \dots ,p_M) &=& 
\sum_{N_1} \sum_{N_2} \cdots \sum_{N_k} 
\delta_N^{N_1 + N_2 + \dots + N_k} 
\\ \nonumber
&\times&
P_{N_1}(p_1,p_2, \dots ,p_{N_1}) \;
P_{N_2}(p_{N_1+1},p_{N_1+2}, \dots ,p_{N_1 + N_2}) 
\\ \nonumber
&\times& \cdots \times
P_{N_k}(p_{N_1+ \dots +N_{k-1}+1}, p_{N_1+ \dots +N_{k-1}+2}, 
\dots ,p_{N_1+ \dots +N_k}) \;.
\ea
One checks that the distribution (\ref{k-source-distri}) is
normalized in agreement with Eq.~(\ref{normal1}). One defines
the single particle distribution analogous to (\ref{singel-distri})
and computes the average single particle momentum at fixed 
multiplicity as
\be
\label{pT-N-k}
\langle p \rangle_{(N)}^k 
=  \frac{1}{{\cal P}_N^k} \frac{1}{N}
\sum_{N_1} \sum_{N_2} \cdots \sum_{N_k} 
\delta_N^{N_1 + N_2 + \dots + N_k} {\cal P}_{N_1}{\cal P}_{N_2}
\cdots {\cal P}_{N_k}
\Big( N_1 \langle p \rangle_{(N_1)} +
N_2 \langle p \rangle_{(N_2)} + \cdots +
N_k \langle p \rangle_{(N_k)} \Big) \;.
\ee

Using the parameterization (\ref{pT-N-corr1}), one finds
\ba
\label{eq-corr1-1}
\langle p \rangle_{(N)}^k 
&=&  \frac{1}{{\cal P}_N^k} \frac{1}{N}
\sum_{N_1} \sum_{N_2} \cdots \sum_{N_k} 
\delta_N^{N_1 + N_2 + \dots + N_k} {\cal P}_{N_1}{\cal P}_{N_2}
\cdots {\cal P}_{N_k}
\\ \nonumber &\times&
\bigg( (N_1 + N_2 + \cdots + N_k)(a + b) 
- b \Big(\frac{N_1^2}{\langle N_1 \rangle} +
\frac{N_2^2}{\langle N_2 \rangle} + \cdots + 
\frac{N_k^2}{\langle N_k \rangle}\Big) \bigg) 
\\ \nonumber
&=&  a + b 
- \frac{b}{\langle N \rangle}\frac{1}{{\cal P}_N^k} \frac{1}{N}
\sum_{N_1} \sum_{N_2} \cdots \sum_{N_k} 
\delta_N^{N_1 + N_2 + \dots + N_k} {\cal P}_{N_1}{\cal P}_{N_2}
\cdots {\cal P}_{N_k}\big(N_1^2+ N_2^2 + \cdots + N_k^2 \big) \;,
\ea
where it has been observed that
$$
N_1 + N_2 + \cdots + N_k = N \;,
$$
$$
\langle N_1 \rangle + \langle N_2 \rangle +\cdots + \langle N_k \rangle 
= \langle N \rangle_k  \;,
$$
$$
{\cal P}_N^k = \sum_{N_1} \sum_{N_2} \cdots \sum_{N_k} 
\delta_N^{N_1 + N_2 + \dots + N_k} {\cal P}_{N_1}{\cal P}_{N_2}
\cdots {\cal P}_{N_k} \;.
$$
The average with the single-source distribution is denoted, as
previously, as $\langle \cdots \rangle$ while the average with 
the $k-$source distribution is $\langle \cdots \rangle_k$. 

Further calculation is performed with the poissonian
multiplicity distribution which is
\ban
{\cal P}_N = \frac{\langle N \rangle^N}{N!}\, e^{-\langle N \rangle } \;.
\ean
The convolution of $k$ identical Poisson distributions of the 
average $\langle N \rangle$ is known to be equal to the Poisson 
distribution of the average $\langle N \rangle_k = k \langle N \rangle$.
Thus,
\be
\label{convo1}
\sum_{N_1} \sum_{N_2} \cdots \sum_{N_k} 
\delta_N^{N_1 + N_2 + \dots + N_k} {\cal P}_{N_1}{\cal P}_{N_2}
\cdots {\cal P}_{N_k}
= \frac{(\langle N \rangle_k)^N}{N!}\, e^{-\langle N \rangle_k} 
={\cal P}_{N}^k\;.
\ee
Using the same technique, which provides Eq.~(\ref{convo1}), one proves
that
\be
\label{convo2}
\sum_{N_1} \sum_{N_2} \cdots \sum_{N_k} 
\delta_N^{N_1 + N_2 + \dots + N_k} {\cal P}_{N_1}{\cal P}_{N_2}
\cdots {\cal P}_{N_k} \, N_1^2
= \Big(\frac{N}{k} + \frac{N(N-1)}{k^2}\Big){\cal P}_{N}^k \;.
\ee
Substituting the formula (\ref{convo2}) into Eq.~(\ref{eq-corr1-1}),
one finds the result 
\be
\label{pT-N-corr1-k}
\langle p \rangle_{(N)}^k = a 
+ b\bigg(1 - \frac{N+k-1}{\langle N \rangle_k}\bigg) \;,
\ee
which gives Eq.~(\ref{pT-N-corr1}) for $k=1$. 
If $\langle N \rangle \gg 1$ or equivalently 
$\langle N \rangle_k \gg k$, the explicit dependence of 
$\langle p \rangle_{(N)}^k$ on $k$ is effectively very weak. For 
the Poisson distribution with $\langle N \rangle_k \gg k \ge 1$, 
the multiplicities $N$ of sizable probabilities do not much differ
than $\langle N \rangle_k$. Consequently, $N+k-1$ can be 
approximated by $N$ in Eq.~(\ref{pT-N-corr1-k}) and we get the 
result analogous to the $k=1$ case (\ref{pT-N-corr1}) that is
\be
\label{pT-N-corr1-k-a}
\langle p \rangle_{(N)}^k \approx a 
+ b\bigg(1 - \frac{N}{\langle N \rangle_k}\bigg) \;.
\ee
Thus, the average momentum at fixed multiplicity is approximately 
independent of the number of sources as it depends only on the 
ratio $N/\langle N \rangle_k$. The correlation strength parameter
$b$ is entirely independent of the source number.

%%%%%%%%%%%%%%%%%%%%%%%%%%%%%%%%%%%%%%%%%%%%%%%%%%%%%%%%%%%%
\section{Varying number of sources}
\label{v-sources}
%%%%%%%%%%%%%%%%%%%%%%%%%%%%%%%%%%%%%%%%%%%%%%%%%%%%%%%%%%%%

In the previous section the number of particle's sources has been 
treated as a fixed number. In a real experimental situation, 
however, the number of sources cannot be fully controlled in 
nucleus-nucleus collisions. The centrality selection is never
perfect but even an exact determination of the participant number
does not guarantee that the number of particle sources is always
the same for a given number of participating nucleons. Thus, 
let us consider what happens when the number $k$ varies event 
by event while the multiplicity of produced particles is kept 
constant. 

When the number of particle sources fluctuates, the momentum 
distribution of $N$ particles coming from $k$ sources,
which is given by Eq.~(\ref{k-source-distri}), has to be 
averaged with the distribution of number of sources 
${\sf P}_k$. The analogue of Eq.~(\ref{pT-N-k}) reads
\ba
\label{pT-N-k-ave}
\langle \langle p \rangle_{(N)}^k \rangle 
&=&  \frac{1}{\langle {\cal P}_N^k\rangle} \frac{1}{N}
\sum_k {\sf P}_k \sum_{N_1} \sum_{N_2} \cdots \sum_{N_k} 
\delta_N^{N_1 + N_2 + \dots + N_k} {\cal P}_{N_1}{\cal P}_{N_2} 
\cdots {\cal P}_{N_k} 
\\[2mm] \nonumber 
&\times& \Big( N_1 \langle p \rangle_{(N_1)} +
N_2 \langle p \rangle_{(N_2)} + \cdots +
N_k \langle p \rangle_{(N_k)} \Big) \;,
\ea
where $\langle {\cal P}_N^k\rangle$ is the multiplicity
distribution averaged over number of sources which equals
\ban
\langle {\cal P}_N^k\rangle =
\sum_k {\sf P}_k \sum_{N_1} \sum_{N_2} \cdots \sum_{N_k} 
\delta_N^{N_1 + N_2 + \dots + N_k} {\cal P}_{N_1}{\cal P}_{N_2} 
\cdots {\cal P}_{N_k} \;.
\ean
Substituting the parameterization (\ref{pT-N-corr1}) into 
Eq.~(\ref{pT-N-k-ave}), one finds
\ba
\label{eq77}
\langle \langle p \rangle_{(N)}^k \rangle 
&=&  \frac{1}{\langle {\cal P}_N^k\rangle} \frac{1}{N}
\sum_k {\sf P}_k \sum_{N_1} \sum_{N_2} \cdots \sum_{N_k} 
\delta_N^{N_1 + N_2 + \dots + N_k} {\cal P}_{N_1}{\cal P}_{N_2} 
\cdots {\cal P}_{N_k} 
\\[2mm] \nonumber 
&\times& 
\bigg((N_1 + N_2 + \cdots + N_k) (a + b)
-b \Big( \frac{N_1^2}{\langle N_1 \rangle} +
\frac{N_2^2}{\langle N_2 \rangle}
+ \cdots + \frac{N_k^2}{\langle N_k \rangle}
\Big) \bigg)
\\[2mm] \nonumber 
&=&  a + b 
- \frac{b}{\langle {\cal P}_N^k \rangle N \langle N \rangle }
\sum_k {\sf P}_k k \sum_{N_1} \sum_{N_2} \cdots \sum_{N_k} 
\delta_N^{N_1 + N_2 + \dots + N_k} {\cal P}_{N_1}{\cal P}_{N_2} 
\cdots {\cal P}_{N_k} \: N_1^2 \;.
\ea
When the single-source multiplicity distribution ${\cal P}_{N}$
is poissonian, one uses the formula (\ref{convo2}) and 
Eq.~(\ref{eq77}) gives
\be
\label{eq78}
\langle \langle p \rangle_{(N)}^k \rangle 
=  a + b \bigg(1
- \frac{1}{\langle N \rangle}
- \frac{N-1}{\langle {\cal P}_N^k \rangle \langle N \rangle}
\sum_k {\sf P}_k \frac{1}{k}\,{\cal P}_{N}^k \bigg)\;.
\ee

Analytic calculations can be easily performed when the number 
of sources is treated not as an integer but as a continuous 
(positive) number distributed according to the Gamma distribution 
\be
\label{gamma}
{\sf P}(k) = \frac{\alpha^{\beta +1}}{\Gamma (\beta + 1)}
k^\beta {\rm e}^{-\alpha k} \;,
\ee
where $\Gamma(x)$ is the gamma function and the parameters 
$\alpha, \; \beta $ obey $\alpha > 0 $ and $\beta > -1 $. 
One easily finds the average and the variance of $k$ as
\ban
\langle k \rangle = \frac{\beta + 1}{\alpha} \;,
\;\;\;\;\;\;\;
\sigma^2 = \frac{\beta + 1}{\alpha^2} \;.
\ean
Substituting the Poisson multiplicity distribution (\ref{convo1})
and the gamma distribution (\ref{gamma}) into Eq.~(\ref{eq78}), 
one finds 
\be
\label{eq79}
\langle \langle p \rangle_{(N)}^k \rangle 
=  a + b \bigg(1
- \frac{1}{\langle N \rangle}
- \frac{\langle N \rangle \sigma^2 + \langle k \rangle}
{\langle N \rangle}
\: \frac{N-1}{(N-1) \sigma^2 + \langle k \rangle^2}\bigg) \;.
\ee
Let us first discuss Eq.~(\ref{eq79}) in two limiting cases. When 
the variance of the source number $\sigma^2$ vanishes, Eq.~(\ref{eq79}) 
reproduces, as expected, the result (\ref{pT-N-corr1-k}) with 
$k = \langle k \rangle$. For $\sigma^2 \rightarrow \infty$, one gets
\be
\label{eq80}
\langle \langle p \rangle_{(N)}^k \rangle 
=  a - \frac{b}{\langle N \rangle} = \langle p \rangle \;.
\ee
Thus, the momentum $\langle \langle p \rangle_{(N)}^k \rangle$ is  
independent of $N$ and it equals the inclusive average - the 
correlation of the transverse momentum and multiplicity 
entirely disappears. 

The correlation of the transverse momentum and multiplicity is 
reduced due to the source number fluctuation. To quantify the effect 
of the correlation reduction, Eq.~(\ref{eq79}) is expanded in the 
powers of $(1 - N/\langle\langle N \rangle_k\rangle)$, where 
$\langle \langle N \rangle_k \rangle \equiv 
\langle k \rangle \langle N\rangle$, and only the linear term 
is kept. Then, Eq.~(\ref{eq79}) changes into
\be
\label{eq81}
\langle \langle p \rangle_{(N)}^k \rangle 
=  a^\prime + b^\prime \bigg(1 - 
\frac{N}{\langle\langle N \rangle_k\rangle}\bigg) \;,
\ee
where $a^\prime \equiv a - 1/\langle N \rangle$ and
\be
\label{corr-reduce}
b^\prime \equiv b \, \frac{\sigma^2 \langle N \rangle \langle k \rangle^3 
+ \langle k \rangle^4}{(\sigma^2 \langle N \rangle \langle k \rangle 
+ \langle k \rangle^2 - \sigma^2)^2} \;.
\ee
When $\sigma^2 \rightarrow 0$, then $b^\prime \rightarrow b$. 
For $\sigma^2 \rightarrow \infty$, $b^\prime \rightarrow 0$.
One also observes that $b^\prime \rightarrow b$ for 
$ \langle k \rangle \rightarrow \infty$. Eq.~(\ref{corr-reduce}) can
be rewritten using the scaled variance 
$\omega \equiv \sigma^2/\langle k \rangle$ as
\ban
b^\prime \equiv b \, \frac{\omega \langle N \rangle + 1}
{(\omega \langle N \rangle + 1 - \omega /\langle k \rangle)^2} \;,
\ean
but it does not eliminate an explicit dependence of $b^\prime$
on $\langle k \rangle$.

%%%%%%%%%%%%%%%%%%%%%%%%%%%%%%%%%%%%%%%%%%%%%%%%%%%%%%%%%%%%
\section{Model predictions}
\label{models}
%%%%%%%%%%%%%%%%%%%%%%%%%%%%%%%%%%%%%%%%%%%%%%%%%%%%%%%%%%%%

We discuss here predictions of the formula (\ref{eq81}), taking 
into account the conditions of the NA49 experiment where the 
particles produced in p--p and Pb--Pb collisions are observed, 
as mentioned in the Introduction, in the forward rapidity window 
$[4.0,5.5]$. The coverage in the azimuthal angle is also incomplete.
The average multiplicity of charged particles in p--p interaction 
is $\langle N \rangle=1.4$ in such an acceptance \cite{Anticic:2003fd}. 

The formula (\ref{eq81}) is first interpreted in the Wounded 
Nucleon Model \cite{Bialas:1976ed} belonging to the transparency
class and, as explained in the Introduction, the source number $k$ 
is identified with the number of wounded nucleons from a projectile 
$N_{\rm proj}$. The number $N_{\rm proj}$ is experimentally determined on the 
event-by-event basis, using the zero-degree calorimeter. Due to 
the finite energy resolution of the calorimeter, the number $N_{\rm proj}$ 
fluctuates with the dispersion identified as $\sigma$.

\begin{figure}
\begin{minipage}{20pc}
\includegraphics[width=19.5pc]{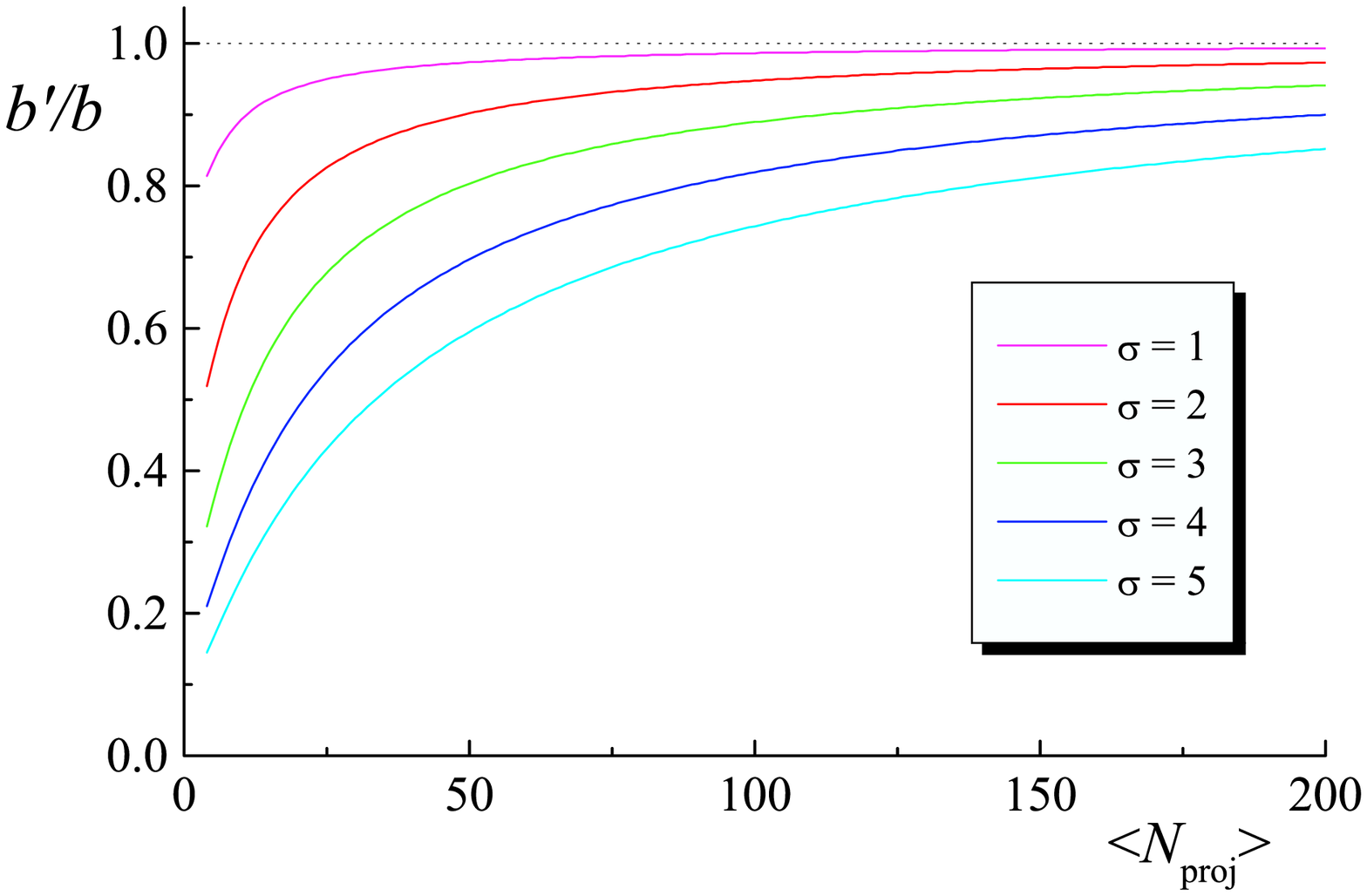}
\vspace{-3.2cm}
\caption{(Color online) The ratio $b^\prime/b$ as a function of the 
average number of projectile wounded nucleons 
$\langle N_{\rm proj} \rangle$ in Pb--Pb collisions for several values 
of the dispersion $\sigma$ of $ N_{\rm proj}$. The most upper line 
corresponds to $\sigma=1$, the lower one to $\sigma=2$, etc.} 
\label{fig1}
\end{minipage}\hspace{2pc}%
\begin{minipage}{20pc}
\includegraphics[width=19.5pc]{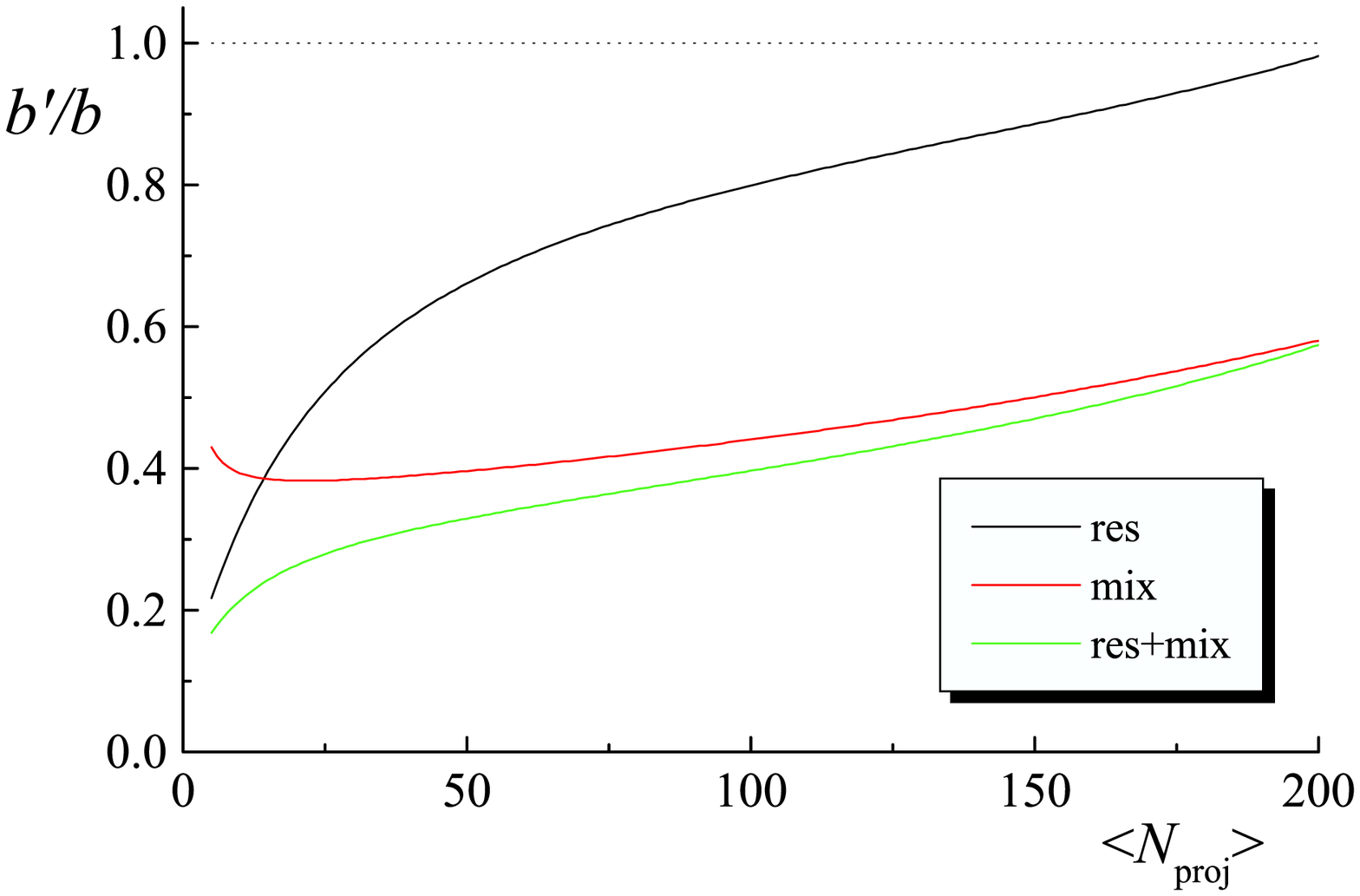}
\vspace{-3.2cm}
\caption{(Color online) $b^\prime/b$ vs. $\langle N_{\rm proj} \rangle$ 
in Pb--Pb collisions. The upper curve (`res') shows the effect of 
calorimeter resolution; the middle one (`mix') represents the mixing 
model; the lower curve (`res+mix') combines the effect of calorimeter
resolution and the mixing-model prediction.} 
\label{fig2}
\end{minipage}
\end{figure}

In Fig.~\ref{fig1} there is shown the ratio $b^\prime/b$, which measures
the correlation reduction, as a function of the average number 
of wounded nucleons $\langle N_{\rm proj} \rangle$ for several values of 
the dispersion $\sigma$ of the wounded nucleon number. As seen, 
the observable correlation strength is strongly reduced for 
the peripheral collisions and it does {\em not} match to the 
p--p point ($b^\prime/b=1$ at $\langle N_{\rm proj} \rangle = 1$), as 
one can naively expect. 

The dispersion of $N_{\rm proj}$ is usually not constant but it changes 
with $\langle N_{\rm proj} \rangle$. The energy resolution of the NA49 
zero-degree calorimeter \cite{Rybczy} which is recalculated into 
the dispersion of projectile participant number $N_{\rm proj}$ is
\be
\label{resol}
\sigma_{\rm proj} (\langle N_{\rm proj} \rangle ) = 0.77 \sqrt{A - \langle N_{\rm proj} \rangle}
-0.28 - 0.032 \: (A - \langle N_{\rm proj} \rangle ) \;,
\ee 
where $A$ is the atomic number of the projectile nucleus. For 
the lead ($A=208$) projectile, the formula (\ref{resol}) gives 
$\sigma$ which varies only between 3.5 and 4.0 for 
$1 < \langle N_{\rm proj} \rangle < 150$, but it drops to 1.6 for 
$\langle N_{\rm proj} \rangle = 200$. The prediction of the formula
(\ref{corr-reduce}) with the dispersion of the wounded nucleon number
given by Eq.~(\ref{resol}) for $A=208$ is represented by the 
most upper curve labeled as `res' in Fig.~\ref{fig2}.

As mentioned in the Introduction, the transparency model, which
assumes that the projectile participants mostly contribute to
the particle production in the forward hemisphere, fails to 
describe \cite{Gazdzicki:2005rr,Konchakovski:2005hq} the large 
multiplicity fluctuations observed in the forward rapidity window 
\cite{Gazdzicki:2004ef,Rybczynski:2004yw}. The mixing models 
\cite{Gazdzicki:2005rr,Konchakovski:2005hq}, where the wounded
nucleons from the projectile and target equally contribute 
to the forward rapidity window\footnote{The analysis performed
at RHIC energies shows that the contribution of a projectile
(target) wound nucleon actually extends far into the backward 
(forward) hemisphere \cite{Bialas:2004su}.}, agree, at least 
qualitatively, with the data \cite{Gazdzicki:2004ef,Rybczynski:2004yw}. 
So, let us consider the model predictions concerning the correlation 
$p_T$~vs.~$N$. The average number of sources $\langle k \rangle$ 
then equals $(\langle N_{\rm proj} \rangle + \langle N_{\rm targ} \rangle )/2$ where 
$N_{\rm targ}$ is the number of wounded nucleons from the target nucleus. 
Since we consider collisions of identical nuclei 
$\langle N_{\rm targ} \rangle = \langle N_{\rm proj} \rangle$, and thus
$\langle k \rangle = \langle N_{\rm proj} \rangle$.

We first neglect the experimental effect of finite
calorimeter resolution, and we assume that $N_{\rm proj}$ is 
precisely determined - $N_{\rm proj}$ does not fluctuate and 
$\langle N_{\rm proj} \rangle = N_{\rm proj}$. The number of particle
sources, which contribute to the forward hemisphere, fluctuates
due to the fluctuations of $N_{\rm targ}$ at fixed $N_{\rm proj}$ and due to
the collision dynamics: a wounded nucleon from the projectile
(target) is assumed to contribute with probability 1/2 to the 
forward (backward) hemisphere. Then, the variance of the particle
sources, which contribute to the forward hemisphere, is
$\sigma^2 = N_{\rm proj}/2 + \sigma_{\rm targ}^2/4$, where $\sigma_{\rm targ}^2$ is 
the variance of $N_{\rm targ}$ at fixed $N_{\rm proj}$. Using the numerical 
results obtained within the Hadron-String Dynamics model 
\cite{Cassing:1999es}, which were used in
\cite{Gazdzicki:2005rr,Konchakovski:2005hq}, $\sigma_{\rm targ}^2$
can be roughly parameterized as
\ban
\sigma_{\rm targ}^2 ( N_{\rm proj} ) = 3.28 \:N_{\rm proj} - 0.0158 N_{\rm proj}^2 \;,
\ean
for the Pb-Pb collisions. As seen, $\sigma_{\rm targ}$ vanishes for $N_{\rm proj}=0$
and $N_{\rm proj}=208$, and it reaches maximum equal 13.1 at $N_{\rm proj}=104$.
Thus, the variance of the particle sources is
\be
\label{mix}
\sigma^2 ( N_{\rm proj} ) = 1.32 \:N_{\rm proj} - 0.004 N_{\rm proj}^2 \;,
\ee 
with maximum equal 10.4 at $N_{\rm proj}=165$. Substituting 
$\langle k \rangle = N_{\rm proj}$ and $\sigma$ given by Eq.~(\ref{mix}) 
into the formula (\ref{corr-reduce}), one finds the result 
shown in Fig.~\ref{fig2} by the middle curve labeled as `mix'. 
As seen, the mixing model leads to a sizeable reduction of the 
correlation $p_T$~vs.~$N$, and the reduction does not much
depened on the collision centrality.

Finally, we have combined the effect of `mixing' with that of
the finite calorimeter resolution. Treating the two sources of 
fluctuations as independent from each other, the variances have 
been summed up. The result is shown in Fig.~\ref{fig2} by the 
lowest curve labeled as `res+mix'.

%%%%%%%%%%%%%%%%%%%%%%%%%%%%%%%%%%%%%%%%%%%%%%%%%%%%%%%%%%%%
\section{Summary and conclusions}
%%%%%%%%%%%%%%%%%%%%%%%%%%%%%%%%%%%%%%%%%%%%%%%%%%%%%%%%%%%%

The correlation $p_T$~vs.~$N$, which is evident in p--p 
collisions, has been studied in the superposition model
of nucleus--nucleus collisions where produced particles come
from $k$ identical and independent sources. At first the number
of sources has been treated as a fixed number, and then the
fluctuations of $k$ have been taken into account. While the 
superposition of particle's sources preserves the correlation 
strength, the correlation is strongly reduced by the source 
number fluctuations. The calculations are fully analytical.

The Wounded Nucleon Model has been used to formulate predictions
for the NA49 measurement of the correlation $p_T$~vs.~$N$ in
Pb--Pb collisions. Since the measurement, which is underway, is 
performed in the forward rapidity window, it has been first 
assumed that only wounded nucleons from the projectile contribute
there. Taking into account a finite energy resolution of the 
zero-degree calorimeter, which allows one to precisely fix the 
number of wounded nucleons from the projectile, the correlation 
has been found to be strongly reduced in the peripheral 
collisions while the reduction in the central collisions is
quite small.

Then, the `mixed' Wounded Nucleon Model 
\cite{Gazdzicki:2005rr,Konchakovski:2005hq} has been considered. 
The model, which appears to be successful in describing
the multiplicity fluctuations at fixed number of the projectile 
wounded nucleons \cite{Gazdzicki:2004ef,Rybczynski:2004yw}, 
assumes, contrary to a natural expectation \cite{Bialas:2004su}, 
that the wounded nucleons from the projectile and target equally 
contribute to the forward rapidity window. In this case, the 
correlation $p_T$~vs.~$N$ seen in the forward rapidity window is 
strongly reduced and the reduction weakly depends on the collision 
centrality.

The correlation $p_T$~vs.~$N$ seems to be responsible 
\cite{Mrowczynski:2004cg} for the transverse momentum 
fluctuations \cite{Anticic:2003fd} as quantified by 
the measure $\Phi(p_T)$ \cite{Gazdzicki:ri}. The data
on the $p_T$~vs.~$N$ correlation combined with the
existing data on the multiplicity 
\cite{Gazdzicki:2004ef,Rybczynski:2004yw} and transverse 
momentum fluctuations \cite{Anticic:2003fd} - all obtained 
in the same experimental conditions - will hopefully allow 
one to formulate a coherent picture of the event-by-event 
fluctuations observed in nucleus--nucleus collisions.

%*************************************************************
\begin{acknowledgements}
                                                                                
I am grateful to Marek Ga\'zdzicki for stimulating discussions
and critical reading of the manuscript. A support by the Virtual 
Institute VI-146 of Helmholtz Gemeinschaft is also acknowledged.
                                                                                
\end{acknowledgements}
%*************************************************************

\end{document}